\RequirePackage{ifpdf}
\ifpdf % We are running pdfTeX in pdf mode
\documentclass[pdftex]{sigma}
\else
\documentclass{sigma}
\fi

\begin{document}

\allowdisplaybreaks

\renewcommand{\PaperNumber}{116}

\FirstPageHeading

\ShortArticleName{Noncommutative Phase Spaces by Coadjoint Orbits Method}

\ArticleName{Noncommutative Phase Spaces\\ by Coadjoint Orbits Method}

\Author{Ancille NGENDAKUMANA~$^\dag$, Joachim NZOTUNGICIMPAYE~$^\ddag$\\ and Leonard TODJIHOUNDE~$^\dag$}

\AuthorNameForHeading{A.~Ngendakumana, J.~Nzotungicimpaye and L.~Todjihounde}

\Address{$^\dag$~Institut de Math\'ematiques et des Sciences Physiques, Porto-Novo, Benin}
\EmailD{\href{mailto:nancille@yahoo.fr}{nancille@yahoo.fr}, \href{mailto:leonardt@imsp-uac.org}{leonardt@imsp-uac.org}}

\Address{$^\ddag$~Kigali Institute of Education, Kigali, Rwanda}
\EmailD{\href{mailto:kimpaye@kie.ac.rw}{kimpaye@kie.ac.rw}}

\ArticleDates{Received May 24, 2011, in f\/inal form December 13, 2011; Published online December 18, 2011}

\Abstract{We introduce noncommutative phase spaces by minimal couplings (usual
one, dual one and their mixing).  We then realize some of them
as coadjoint orbits of the aniso\-tropic Newton--Hooke groups in two- and three-dimensional spaces. Through these constructions the positions and the momenta of the
phase spaces do not commute due to the presence of a magnetic f\/ield and a dual magnetic f\/ield.}


\Classification{22E60; 22E70; 37J15; 53D05; 53D17}

\section{Introduction}

Noncommutative phase spaces provide mathematical backgrounds for the study of magnetic f\/ields in physics.
 Noncommutativity appeared in nonrelativistic mechanics f\/irst in the work of
 Peierls \cite{peierls} on the diamagnetism of conduction electrons.
In relativistic quantum mechanics, noncommutativity was f\/irst examined in 1947
 by Snyder~\cite {snyder}.  During the last $15$ years, noncommutative mechanics has been an important subject which attracted quite a lot of attention
(see, e.g.,~\cite{horvathy11, horvathy12, horvathy13, grigore, horvathy2,vanhecke2}).

Noncommutative phase space is def\/ined as a space on which coordinates satisfy the commutation relations:
 \begin{gather*}
 \{ q^i,q^j \}=G^{ij} ,\qquad   \{ q^i,p_j \}=\delta^{i}_j , \qquad \{ p_i,p_j \}= F_{ij}, % \label{poisson1}
\end{gather*}
where $\delta^{i}_j$ is a unit matrix, whereas $G^{ij}$ and $F_{ij}$ are functions of positions and momenta. The physical dimensions of $G^{ij}$ and $F_{ij}$ are respectively $M^{-1}T$ and $MT^{-1}$, where $M$ represents a~mass while $T$ represents a time.

The aim of this paper is to introduce the construction of noncommutative spaces by using
dif\/ferent minimal couplings and the realization of some of them as coadjoint orbits~\cite{kirillov,kostant,souriau}.  The maximal coadjoint orbits of the anisotropic Newton--Hooke groups in two dimensions and in three dimensions are shown to be models of noncommutative spaces.

The paper is organized as follows. In Section~\ref{section2} noncommutative phase spaces are introduced by generalizing the usual
Hamiltonian equations to the cases where a magnetic f\/ield and a dual magnetic f\/ield are present. Section~\ref{section3} is devoted to the study of
planar mechanics in the following three situations:
\begin{itemize} \itemsep=0pt
\item when a massive charged particle is in an electromagnetic f\/ield,
\item when a massless spring is in a
dual magnetic f\/ield,
\item when a pendulum is in an electromagnetic f\/ield and in a dual one.
\end{itemize}

 It is shown in this paper that under the presence of these f\/ields
 \begin{itemize}\itemsep=0pt
 \item the massive charged particle acquires an oscillatory motion with a certain frequency,
 \item the massless spring acquires a mass,
\item the pendulum appears like two synchronized oscillators.
 \end{itemize}
The second and third results mentioned above are quite new. In Section~\ref{section4} we construct, for the f\/irst time, the coadjoint orbits
 of the anisotropic Newton--Hooke groups in dimension two and dimension three. Then, we obtain noncommutative phase spaces in presence of a magnetic
 f\/ield and a dual magnetic f\/ield.

\section{Noncommutative phase spaces}\label{section2}

In this paragraph we recall Hamiltonian mechanics in both Darboux's coordinates (Section~\ref{section2.1}) and noncommutative
coordinates (Section~\ref{section2.2}), the noncommutativity coming from the pre\-sen\-ce of two f\/ields~$F_{ij}$ and~$G^{ij}$.  We will distinguish three cases of
noncommutative coordinates corresponding to the presence of the magnetic f\/ield only,
of the dual magnetic f\/ield only and of the both f\/ields simultaneously.

\subsection{Commutative coordinates}\label{section2.1}

It is known that a symplectic manifold is a $2n$-dimensional manifold equipped with a closed nondegenerate $2$-form $\sigma$.
If $\sigma_{ab}$ are the matrix elements of the matrix representing the symplectic form $\sigma$ and if $\sigma^{ab}$ are solutions of
$\sigma_{bc}\sigma^{ca}=\delta^a_b$, then a Poisson bracket of  two functions $f$ and $g$ belonging to $C^{\infty}(V,{\mathbb R})$ is given by
\begin{gather}\label{pb}
\{f,g\}=\sigma^{ab}\frac{\partial f}{\partial z^a}\frac{\partial g}{\partial z^b}.
\end{gather}
The space  $C^{\infty}(V,{\mathbb R})$ endowed with the Poisson bracket given by (\ref{pb}) is an inf\/inite Lie algebra~\cite{abraham}.
If $z^a=(p_i,q^i)$ are the canonical coordinates (Darboux's coordinates) on~$V$,
the symplectic form on $V$ is $\sigma=dp_i\wedge dq^i$, that means there is no coupling to a gauge f\/ield, and the Poisson bracket~(\ref{pb}) becomes
\begin{gather*}
\{f,g\}=\frac{\partial f}{\partial p_i}\frac{\partial
g}{\partial q^i}-\frac{\partial f}{\partial q^i}\frac{\partial
g}{\partial p_i}.
\end{gather*}
It follows that
\begin{gather*}
\{p_k,p_i\}=0 ,\qquad \{p_k,q^i\}=\delta^i_k , \qquad \{q^k,q^i\}=0.
\end{gather*}
That means the momenta $p_i$  as well as the positions
$q^i$ are commutative.

It is also known that if $X_f$ is the Hamiltonian vector f\/ield associated to $f$, then
$X_f(g)=\{f,g\}$ and the evolution equations under the f\/low $\Phi_{\exp(sX_f)}$ on $V$ generated by $X_f$ are
\begin{gather*}
\frac{dz^a}{ds}=X_f(z^a),
\end{gather*}
which are exactly the usual Hamiltonian equations when $f$ is the energy.

 Let us now introduce noncommutative coordinates by coupling the momentum $p_i$ with a~magnetic potential $A_i$ and the position $q^i$ with a
potential $A^{*i}$.

\subsection{Noncommutative coordinates}\label{section2.2}

Let us consider the change of coordinates
\begin{gather}\label{newcoordinates}
\pi_i=p_i-\frac{1}{2}F_{ik}q^k ,\qquad x^i=q^i-\frac{1}{2}p_kG^{ki}.
\end{gather}
The matrix form of (\ref{newcoordinates}) is
\begin{gather*}
\begin{pmatrix}
\pi_i\\x^i
\end{pmatrix} =\begin{pmatrix}
\delta^k_i&-\frac{1}{2}F_{ik}\vspace{1mm}\\-\frac{1}{2}G^{ik}&\delta^i_k 
\end{pmatrix} \begin{pmatrix}
p_k\\q^k
\end{pmatrix}.
\end{gather*}
As
\begin{gather*}
\begin{pmatrix}
\delta^k_i&-\frac{1}{2}F_{ik}\vspace{1mm}\\-\frac{1}{2}G^{ik}&\delta^i_k
\end{pmatrix}
=\begin{pmatrix}
\delta^j_i&-\frac{1}{2}F_{ij}\vspace{1mm}\\0^{ij}&\delta^i_j
\end{pmatrix} \begin{pmatrix}
\delta^s_j-\frac{1}{4}F_{jm}G^{ms}&0_{js}\vspace{1mm}\\0^{js}&\delta^j_s
\end{pmatrix} \begin{pmatrix}
\delta^k_s&0^{sk}\vspace{1mm}\\-\frac{1}{2}G^{sk}&\delta^s_k
\end{pmatrix},
\end{gather*}
 the transformation (\ref{newcoordinates}) is a change of coordinates if $\det\big(\delta^s_j-\frac{1}{4}F_{jm}G^{ms}\big)\neq 0 $. It follows that  \cite[equation~(5)]{wei}
\begin{gather}\label{newpoisson}
\{\pi_i,\pi_k\}=F_{ik} ,\qquad \{\pi_i,x^k\}=\delta^{k}_i ,\qquad \{x^i,x^k\}=G^{ik},
\end{gather}
i.e.\ the new momenta  as well as the new conf\/iguration
coordinates are noncommutative. The Jacobi identity implies that the $F_{ij}$'s depend only on positions, that the $G^{ij}$'s depend only on momenta
and that the two $2$-forms $\sigma_1=F_{ij}(x)dx^i\wedge dx^j$ and
$\sigma_2=G^{ij}(\pi)d\pi_i\wedge d\pi_j$ are closed.

Let the Poisson brackets of two functions $f$, $g $ in the new coordinates be given by
\begin{gather*}
\{f,g\}_{\rm new}=\frac{\partial f}{\partial \pi_i}\frac{\partial
g}{\partial x^i}-\frac{\partial f}{\partial x^i}\frac{\partial
g}{\partial \pi_i}=Y_f(g).
\end{gather*}
It follows that
\begin{gather*}
Y_H=X_H+G^{ij}\frac{\partial H}{\partial q^i}\frac{\partial }{\partial
q^j}+F_{ij}\frac{\partial H}{\partial p_i}\frac{\partial }{\partial
p_j}.
\end{gather*}
The derivative of any function $f$ with respect to time $t$, in terms of $F$ and $G$, is then given by
\begin{gather}\label{genequamotion}
\frac{df}{dt}=X_H(f)+G^{ij}\frac{\partial H}{\partial q^i}\frac{\partial f }{\partial
q^j}+F_{ij}\frac{\partial H}{\partial p_i}\frac{\partial f }{\partial
p_j},
\end{gather}
and the equations of motion are then given by
\begin{gather*}
\frac{dq^k}{dt}=\frac{\partial H}{\partial p_k}+G^{ki}\frac{\partial
H}{\partial q^i} ,\qquad \frac{dp_k}{dt}=-\frac{\partial H}{\partial
q^k}+F_{ik}\frac{\partial H}{\partial p_i}.
\end{gather*}
 If for example
\begin{gather*}
H=\frac{\delta^{ij}p_ip_j}{2m}+V
\end{gather*}
is the Hamiltonian with the potential energy $V$
depending only on the conf\/iguration coordina\-tes~$q^i$, the equations of motion are then
\begin{gather*}
\frac{dq^k}{dt}=\frac{p^k}{m}+G^{ki}\frac{\partial V}{\partial
q^i} ,\qquad \frac{dp_k}{dt}=-\frac{\partial V}{\partial
q^k}+F_{ik}\frac{p^i}{m}.
\end{gather*}
They are equivalent to the modif\/ied Newton's second law~\cite{horvathy13,romero,wei}
\begin{gather*}
m\frac{d^2q^k}{dt^2}=-\frac{\partial V}{\partial
q^k}+F_{ik}\frac{p^i}{m}+m G^{ki}\frac{d}{dt}\left(\frac{\partial
V}{\partial q^i}\right).
\end{gather*}
This means that the noncommutativity of the momenta implies that the particle is accelerated and is not free even if the potential
$V$ vanishes identically.

\section{Couplings in planar mechanics}\label{section3}

In this section we construct explicitly these noncommutative phase spaces by
introducing couplings. We start with the usual coupling of momentum
with a magnetic potential. Then, we introduce the coupling of
 position with a dual potential and f\/inish with a mixing model.

\subsection{Coupling of momentum with a magnetic f\/ield}\label{section3.1}

\subsubsection{Commutative coordinates}\label{section3.1.1}

Consider a four-dimensional phase space (a cotangent space to a plane) equipped with the~Dar\-boux's coordinates $(p_i,q^i$).
This means that the momenta as well as the positions commute. Consider also an electron with mass $m$ and an
electric charge $e$, moving on a plane with the electromagnetic potential $A_{\mu}=\big(A_i=-\frac{1}{2}B\epsilon_{ik}q^k,\phi=E_iq^i\big)$
 where a symmetric gauge has been chosen, $\vec{E}$ being an electric f\/ield while $\vec{B}$ is a magnetic f\/ield perpendicular to the plane.
It is known that the dynamics of the particle is governed by the
Hamiltonian
\begin{gather}\label{hamiltonianelectron}
H=\frac{\vec{p}\,{}^{2}}{2m}-e\phi
\end{gather}
and that the equation of motion is
\begin{gather*}
m\frac{d^2\vec{q}}{dt^2}=e\vec{E},
\end{gather*}
where the right hand side is the electric force.

\subsubsection{Noncommutative coordinates}\label{section3.1.2}

From the classical electromagnetism, it is known that the coupling of the momentum with the magnetic
potential is given by the relations
\begin{gather}\label{magneticcoupling1}
 \pi_i=p_i+\frac{eB}{2} \epsilon_{ik}q^k ,\qquad x^i=q^i.
\end{gather}
The coordinates $\pi_i$ and $x^i$ are such that
\begin{gather*}
\{x^i,x^k\}=0 ,\qquad \{\pi_i,x^k\}=\delta^k_i , \qquad \{\pi_i,\pi_k\}=-eB\epsilon_{ik}.
\end{gather*}
In the presence of an electromagnetic f\/ield, the momenta are
noncommutative while the positions are commutative.  Using the equation
(\ref{newpoisson}), we have
\begin{gather}\label{magnetic}
F_{ij}=-eB\epsilon_{ij} , \qquad G^{ij}=0^{ij}.
\end{gather}
Use of (\ref{hamiltonianelectron}) and (\ref{magnetic}) into (\ref{genequamotion}) gives rise the equations of motion
\begin{gather}\label{newton1}
m\frac{d^2\vec{q}}{dt^2}=e\left(\vec{E}+\frac{\vec{p}}{m}\times \vec{B}\right),
\end{gather}
where the right hand side represents the Lorentz force.
 Moreover, in noncommutative coordinates, the Hamiltonian (\ref{hamiltonianelectron}) becomes
\begin{gather*}
H=\frac{\vec{\pi}\,{}^{2}}{2m}-e\vec{E}\cdot \vec{x}+\frac{m\omega^2\vec{x}\,{} ^{ 2}}{2}+\vec{\omega}
\cdot \vec{L},
\end{gather*}
where $\omega$ is the {\it cyclotron frequency},
$\vec{L}=\vec{x}\times \vec{p}$ is the orbital angular momentum and
\begin{gather*}
\vec{\omega}=\frac{eB}{2m}\vec{n}
\end{gather*}
with $\vec{n}$ the unit vector in the direction perpendicular to
the plane.
In the presence of a magnetic f\/ield, the massive particle has become an oscillator with frequency $\omega$
given above and then the equation of motion is
\begin{gather}\label{newton2}
m\frac{d^2\vec{x}}{dt^2}=e\left(\vec{E}+\frac{\vec{\pi}}{m}\times\vec{B}\right),
\end{gather}
where we recognize again the Lorentz force
$\vec{f}_{\rm Lorentz}=e\vec{E}+e\frac{\vec{\pi}}{m}\times \vec{B}$.
Note that the relations~(\ref{newton1}) and~(\ref{newton2}) have the
same form. The Newton's equations are then covariant under the
coupling~(\ref{magneticcoupling1}).

In the next two subsections, we present quite new theories associated with an unusual coupling of
position with a dual magnetic f\/ield.

\subsection{Coupling of position with a dual f\/ield}\label{section3.2}

\subsubsection{Commutative coordinates}\label{section3.2.1}

 Consider a massless spring with $k$ as a Hooke's constant and a dual charge $e^*$ in a dual magnetic f\/ield $B^*$.  Suppose that the dynamics of the
spring is governed by the Hamiltonian
\begin{gather}\label{hamdual}
H=k\frac{\vec{q}\,{}^2}{2}-e^*\vec{p}\cdot \vec{E^*},
\end{gather}
where we have used the symmetric gauge.
  Moreover the analogue of the
Newton's second equation is
\begin{gather*}
\frac{1}{k}\frac{d^2\vec{p}}{dt^2}=e^*\vec{E^*},
\end{gather*}
where $\frac{d^2\vec{p}}{dt^2}$ is a yank, i.e.\ the second derivatives of momentum
with respect to the time variable~$t$,  $e^*\vec{E^*}$ is a velocity while $e\vec{E}$  is a
force as in the previous subsection.

\subsubsection{Noncommutative coordinates}\label{section3.2.2}

Let us consider the
coupling of the position with the dual potential $A^{*i}$ depending on the momenta $p_i$,
\begin{gather}\label{coupling2}
  \pi_i=p_i ,\qquad x^i=q^i+\frac{e^*B^*}{2}p_k\epsilon^{ki}.
\end{gather}
In this case, the Poisson brackets become
\begin{gather*}
\{x^i,x^j\}=-e^*B^*\epsilon^{ij} ,\qquad \{p_k,x^i\}=\delta^i_k ,\qquad \{p_k,p_i\}=0.
\end{gather*}
Therefore, in the presence of the dual f\/ield, positions do not commute
while the momenta commute. Then
\begin{gather}\label{dualmagnetic}
F_{ij}=o_{ij} ,\qquad G^{ij}=-e^*B^*\epsilon^{ij}.
\end{gather}
Use (\ref{hamdual}) and (\ref{dualmagnetic}) into (\ref{genequamotion}) gives rise to the Newton's analogue equations are then
\begin{gather}\label{newton3}
\frac{1}{k}\frac{d^2\vec{p}}{dt^2}=e^*\vec{E^*}+e^*k\vec{q}\times
\vec{B^*},
\end{gather}
the right hand side being a velocity.
In noncommutative coordinates the Hamiltonian is
\begin{gather*}
H=\frac{k\vec{x}\,{}^{2}}{2}-e^*\vec{\pi}\cdot \vec{E^*}+\frac{\vec{\pi}\,{}^{2}}{2m_s}-\vec{\omega}\cdot \vec{L},
\end{gather*}
where the spring mass $m_s$ is def\/ined by
\begin{gather*}
\frac{1}{m_s}=k\frac{e^{*2}B^{*2}}{4},
\end{gather*}
while the vector $\vec{\omega}$ is given by
\begin{gather*}
\vec{\omega}=k\frac{e^*B^*}{2}\vec{n}.
\end{gather*}
The Hooke's constant $k$ can be written as
\begin{gather}\label{hookeconstant}
k=m_s\omega^{2}.
\end{gather}
In the presence of the dual f\/ield, the spring then acquires a mass
$m_s$ and the equations of motion are given by
\begin{gather}\label{newton4}
\frac{1}{k}\frac{d^2\vec{\pi}}{dt^2}=e^*\vec{E^*}+e^*k\vec{x}\times \vec{B^*}
.
\end{gather}
The vector $\vec{f^*}=e^*(\vec{E^*}+k\vec{x}\times\vec{B^*})$ can
be considered as a dual Lorentz force with the dimension of velocity. It represents for the spring
what the Lorentz force represents for a charged  particle. Here also the
coupling (\ref{coupling2}) preserves the covariance of the Newton's analogue equations.
Comparing (\ref{newton3}) and (\ref{newton4}) and using~(\ref{hookeconstant}) into~(\ref{newton4}) we can conclude that
$e^*\omega^2(\vec{E^*}+k\vec{x}\times\vec{B^*})$  is a kind of
jerk \cite{nzo2}.

\subsubsection{Coupling with a magnetic f\/ield and with a dual f\/ield}\label{section3.2.3}

 Now consider the case of a massive pendulum with mass $m$ and
Hooke's constant $k$ under the action of an electromagnetic potential $A_{\mu}=(A_i,\phi)$
 and  a dual electromagnetic potential $A^*_{\mu}=(A^*_i,\phi^*)$ with $A_i=-\frac{1}{2}B\epsilon_{ik}q^k$, $\phi=E_iq^i$,
$A^*_i=-\frac{1}{2}B^* p_k\epsilon_{ki}$ and  $\phi^*=p_iE^*_i$, where $\vec{E}$ is an electric f\/ield
and $\vec{E}^*$ its dual f\/ield while $\vec{B}$ is a magnetic f\/ield and $\vec{B}^*$ its dual f\/ield.

The corresponding motion is governed by the Hamiltonian
 \begin{gather*}%\label{hammixt}
H=\frac{\vec{p}\,{} ^2}{2m}+\frac{k\vec{q}\,{}^{2}}{2}-e\phi-e^*\phi^*.
\end{gather*}
Let
\begin{gather*}%\label{coupling5}
x^i=q^i+\frac{e^*B^*}{2}p_k\epsilon^{ki} ,\qquad \pi_i=p_i+\frac{eB}{2}\epsilon_{ik}q^k
\end{gather*}
be the minimal coupling in the symmetric gauge; that is
\begin{gather*}%\label{mixt}
G^{ij} = -e^*B^*\epsilon^{ij}  ,\qquad F_{ij} =- eB \epsilon_{ij}.
\end{gather*}
We assume that the cyclotron frequency acquired by the massive charged particle is equal to the
frequency of the massless spring:
\begin{gather*}
\frac{eB}{2}=m\omega  , \qquad  \frac{e^*B^*}{2}=\frac{1}{m_s\omega},
\end{gather*}
where $m_s$ is the acquired mass by the spring while
\begin{gather*}
\mu=\frac{m\cdot m_s}{m+m_s}
\end{gather*}
is the reduced mass of the two synchronized massive oscillators. It follows that
\begin{gather}\label{noncommutativecoordinates}
\{x^i,x^j\}=-e^*B^*\epsilon^{ij} ,\qquad \{\pi_k,x^i\}=\gamma\delta^i_k ,\qquad \{\pi_k,\pi_i\}=-eB\epsilon_{ki}
\end{gather}
with $\gamma=1+\frac{m}{m_s}$ and $m=\mu\gamma$.

In the presence of the two kind of f\/ield, the positions  as well as the momenta do not
commute.
The Hamiltonian in noncommutative coordinates is written as
\begin{gather*}
H=\frac{\vec{\pi}\,{} ^{ 2}}{2\mu}+\frac{M
\omega^2\vec{x}\,{}^{ 2}}{2}-e\phi -e^* \phi^*,
\end{gather*}
where $M = m + m_s$ is the total mass, $ \phi=\vec{E}\cdot \vec{x}+\vec{n}\cdot \vec{E}\times
\frac{\vec{\pi}}{m_s\omega} $   and $\phi^*=\vec{\pi}\cdot \vec{E^*}+\vec{n}\cdot m\omega
\vec{x}\times \vec{E^*}$. Note that $M=m_s\gamma$.

The motion's equations in noncommutative coordinates are then
\begin{gather*}
\frac{d\vec{x}}{dt}=\frac{\vec{\pi}}{\mu}+e^*\big[\gamma\vec{E^*}+k\vec{x}\times\vec{B^*}
-e\vec{B^*}\times \vec{E}\big],
\\
\frac{d\vec{\pi}}{dt}=-k\gamma\vec{x}+e\left[\gamma\vec{E}+\frac{\vec{\pi}}{m}\times \vec{B}
-e^*\vec{B}\times \vec{E^*}\right],
\end{gather*}
where the Hooke's constant $k$ is given by (\ref{hookeconstant}).

 If the
mass $m$ of the particle is very smaller than the mass~$m_s$
acquired by the spring, i.e.\ $m\ll m_s$, then $\gamma$ becomes $1$ and $\mu=m\ll M=m_s$. In that limit, the brackets~(\ref{noncommutativecoordinates}) become
\begin{gather*}
\{x^i,x^j\}=-e^*B^*\epsilon^{ij} ,\qquad \{\pi_k,x^i\}=\delta^i_k ,\qquad \{\pi_k,\pi_i\}=-eB\epsilon_{ki},
\end{gather*}
the Hamiltonian becomes
\begin{gather*}
H=\frac{\vec{\pi}\,{}^{2}}{2m}+\frac{m_s
\omega^2\vec{x}\,{}^{ 2}}{2}-e\left[\vec{E}\cdot \vec{x}+\vec{n}\cdot \vec{E}\times
\frac{\vec{\pi}}{m_s\omega}\right]-e^*\big[\vec{\pi}\cdot \vec{E^*}+\vec{n}\cdot m\omega
\vec{x}\times \vec{E^*}\big]
\end{gather*}
and the equations of motion  are given by
\begin{gather*}
\frac{d\vec{x}}{dt}=\frac{\vec{\pi}}{m}+e^*\big[\vec{E^*}+k\vec{x}\times\vec{B^*}
-e\vec{B^*}\times \vec{E}\big],
\\
\frac{d\vec{\pi}}{dt}=-k\vec{x}+e\left[\vec{E}+\frac{\vec{\pi}}{m} \times \vec{B}
-e^*\vec{B}\times \vec{E^*}\right].
\end{gather*}
The velocity $ee^*\vec{B^*}\times \vec{E}$ and the force
$ee^*\vec{E^*}\times \vec{B}$ result from the coexistence of the two
f\/ields.

\section[Noncommutative phase spaces as coadjoint orbits of anisotropic Newton-Hooke groups]{Noncommutative phase spaces as coadjoint orbits\\ of anisotropic Newton--Hooke groups}\label{section4}

It is well known that the dual of a Lie algebra has a natural Poisson structure whose symplectic leaves are the coadjoint orbits.
These orbits will provide naturally noncommutative phase spaces.

 In this section, we use orbit construction to realize noncommutative phase spaces on
the~ani\-so\-tropic New\-ton--Hooke groups in two- and three-dimensional spaces, the anisotropic
New\-ton--Hooke groups $ANH_{\pm}$ being Newton--Hooke groups~$NH_{\pm}$
 without the rotation parameters~\cite{derome}. Their Lie algebras have the
structures
\begin{gather*}
[K_i,E]=P_i , \qquad [P_i,E]=\pm \omega^2K_i ,\qquad i=1,2,\dots,n,
\end{gather*}
where
\begin{gather*}
 \vec{K}=\frac{\partial}{\partial
\vec{v}} ,\qquad \vec{P}=\frac{\partial}{\partial
\vec{x}} , \qquad E=\frac{\partial}{\partial
t}+\vec{v}\cdot \frac{\partial}{\partial \vec{x}}\pm
\omega^2\vec{x}\cdot \frac{\partial}{\partial \vec{v}}.
\end{gather*}
Standard methods \cite{hamermesh, kirillov, kostant, nzo1}
show that the structure of the central extensions of the Lie algebras
${\cal{ANH}}_{\pm}$ is
\begin{itemize}\itemsep=0pt
\item in one-dimensional space
\begin{gather*}
[K,E]=P ,\qquad [P,E]=\pm \omega^2K ,\qquad [K,P]=M,
\end{gather*}
\item  in two-dimensional spaces
\begin{gather*}
[K_i,K_j] = \frac{1}{c^2}J_3\epsilon_{ij} , \qquad [K_i,E]=P_i ,\qquad [K_i,P_j]=M\delta_{ij},\\
[P_i,P_j] = \pm \frac{1}{r^2}J_3\epsilon_{ij} ,\qquad [P_i,E]=\pm
\omega^2K_i,
\end{gather*}
\item in three-dimensional spaces
\begin{gather*}
[K_i,K_j] = \frac{1}{c^2}J_k\epsilon^k_{ij} ,\qquad [K_i,E]=P_i ,\qquad [K_i,P_j]=M\delta_{ij},\\
[P_i,P_j] = \pm\frac{1}{r^2}J_k\epsilon^k_{ij} ,\qquad [P_i,E]=\pm \omega^2K_i,
\end{gather*}
\end{itemize}
where $r$ is a constant with the dimension of length, $c$ is a
constant with the dimension of speed while~$J_{k}$ is a rotation parameter around the $k^{\rm th}$ axis.

 \subsection{One-dimensional space case}\label{section4.1}

 In this case $m$ is a trivial invariant. The other invariant, the solution of the Kirillov's system, is
\begin{gather*}
U=e-\frac{p^2}{2m}\pm \frac{m\omega^2q^2}{2},
\end{gather*}
where $q=\frac{k}{m}$.  We denote the two-dimensional orbit by
${\cal{O}}_{(m,U)}$. It is not interesting
for our study because there are one momentum and one position. Note that the symplectic realizations of $ANH_{-}$ and $ANH_{+}$ are respectively given by
\begin{gather*}
 L_{(v,x,t)}(p,q) =  \Big(p\cos (\omega t) - m\omega q\sin (\omega t) - mv\cos (\omega t) , \\
\hphantom{L_{(v,x,t)}(p,q) =  \Big(}{}\frac{p}{m\omega}\sin (\omega t) + (q + x) \cos (\omega t ) -\frac{v}{\omega}\sin (\omega t)\Big)
\end{gather*}
and
\begin{gather*}
L_{(v,x,t)}(p,q)= \Big(p\cosh (\omega t) + m\omega q\sinh (\omega t) -m(v\cosh (\omega t) -\omega x\sinh (\omega t)) ,  \\
\hphantom{L_{(v,x,t)}(p,q)= \Big(}
\frac{p}{m\omega}\sinh (\omega t) + (q+x)\cosh (\omega  t) -\frac{v}{\omega}\sinh (\omega t)\Big).
\end{gather*}
Let
 $(p(t),q(t))=L_{(0,0,t)}(p,q) $, it follows that
\begin{gather*}
 p(t) = p\cos (\omega t)  -m\omega q\sin (\omega t)  ,\qquad  q(t) =\frac{p}{m\omega} \sin (\omega t) + q\cos (\omega t)
\end{gather*}
 for $ANH_{-}$ and
\begin{gather*}
 p(t) =  p\cosh (\omega t) + m\omega q\sinh (\omega t )  ,\qquad  q(t) = \frac{p}{m\omega} \sinh (\omega t)+ q\cosh(\omega t)
\end{gather*}
for $ANH_{+}$. The equations of motion are then given by
\begin{gather*}
 \frac{dp}{dt} = \pm m\omega^{2}q  , \qquad  \frac{dq}{dt} = \frac{p}{m}
\end{gather*}
for $ANH_{\pm}$ or equivalently  $ \frac{d^{2}q}{dt^{2}} = \pm
\omega^{2}q $; which is a second order dif\/ferential equation whose
solutions are trigonometric functions for $ANH_{-}$ case and
hyperbolic ones in $ANH_{+}$ case.  It is for this reason that
$ANH_{-}$ describes a universe in oscillation while $ANH_{+}$
describes a universe in expansion.

\subsection{Two-dimensional spaces case}\label{section4.2}

Let $mM^*+hJ^{*3}+k_iK^{*i}+p_iP^{*i}+eE^*$ be the general element of
the dual of the central extended Lie algebra. Then $m$ and $h$ are
trivial invariants under the coadjoint action of $ANH_{\pm}$ in two-dimensional spaces.  The other invariant, the solution of the Kirillov's
system, is explicitly given~by
\begin{gather*}
U=e-\frac{\vec{p}\,{}^2}{2\mu_e}\pm\frac{\mu_e\omega^2\vec{q}\,{}^2}{2}
\end{gather*}
with
\begin{gather}\label{effectivemass}
\mu_e=m\pm \frac{h}{\omega r^2} , \qquad \vec{q}=\frac{\vec{k}}{\mu_e},
\end{gather}
where $h\omega_0=mc^2$ denotes the wave-particle duality, $\mu_e$ is an ef\/fective mass.  The restriction
of the Kirillov's matrix on the orbit is given by
\begin{gather*}
\Omega=\begin{pmatrix}
0&\frac{h}{c^2}&m&0\vspace{1mm}\\-\frac{h}{c^2}&0&0&m\vspace{1mm}\\-m&0&0&\pm\frac{h}{r^2}\vspace{1mm}\\0&-m&\mp\frac{h}{r^2} &0
\end{pmatrix}.
\end{gather*}
By using relations (\ref{effectivemass}), the duality wave-particle and the equality $c=\omega r$,
we obtain that the Poisson brackets of two functions def\/ined on the orbit are given by
\begin{gather*}
\{H,f\}=\frac{\partial H}{\partial p_i}\frac{\partial f}{\partial
q^i}-\frac{\partial H}{\partial q^i}\frac{\partial f}{\partial
p_i}+G^{ij}\frac{\partial H}{\partial q^i}\frac{\partial f}{\partial
q^j}+F_{ij}\frac{\partial H}{\partial p_i}\frac{\partial f}{\partial
p_j}, \qquad i,j=1,2
\end{gather*}
with
\begin{gather*}
 G^{ij}=-\frac{\epsilon^{ij}}{m\omega_0} ,\qquad  F_{ij}=-(m-\mu_{e})\omega\epsilon_{ij}.
\end{gather*}
 It follows that the magnetic f\/ield $B$ and its dual f\/ield $B^*$ are such that
\begin{gather*}
e^*B^*=\frac{1}{m\omega_0} ,\qquad eB=(m-\mu_e)\omega.
\end{gather*}
The ef\/fective mass is then given in terms of the magnetic f\/ield
as
\begin{gather*}%\label{effectivemass1}
\mu_e=m-\frac{eB}{\omega}.
\end{gather*}
The Hamilton's equations are then
\begin{gather*}%\label{hamiltonequationtwofields}
\frac{d\pi_i}{dt}=-\frac{\partial H}{\partial q^i}\pm
(m-\mu_e)\omega\epsilon_{ik}\frac{\partial H}{\partial
p_k} ,\qquad \frac{dx^i}{dt}=\frac{\partial H}{\partial
p_i}+\frac{\epsilon^{ik}}{2m\omega_0}\frac{\partial H}{\partial q^k}.
\end{gather*}
The inverse of $\Omega$ is
\begin{gather*}
\Omega^{-1}=\begin{pmatrix}
0& \pm
\frac{\omega}{\mu_e}&-\frac{1}{\mu_e}&0\vspace{1mm}\\\mp\frac{\omega}{\mu_e
}&0&0&-\frac{1}{\mu_e}\vspace{1mm}\\\frac{1}{\mu_e}&0&0&\frac{1}{\mu_e\omega_0}\vspace{1mm} \\0&\frac{1}{\mu_e}&-\frac{1}{\mu_e \omega_0}&0
\end{pmatrix},
\end{gather*}
where  we have used the wave-particle duality and (\ref{effectivemass}).
Finally the orbit is equipped with the symplectic form
\begin{gather*}
\sigma=dp_i\wedge
dq^i+\frac{1}{\mu_e\omega_0}\epsilon^{ij}dp_i\wedge dp_j\pm
\mu_e\omega \epsilon_{ij} dq^i\wedge dq^j.
\end{gather*}
We observe that with the anisotropic Newton--Hooke group in two-dimensional spaces, the obtained phase spaces are completely noncommutative
while the phase space obtained with the Galilei group in~\cite{horvathy11} is only partially noncommutative. This is due to the dif\/ference in the structure of their extended Lie algebras. In the Newton--Hooke case space translations as well pure Newton--Hooke transformations do not commute while only pure Galilei transformations do not commute in the Galilei case. Note that the nontrivial Lie brackets of the extended Newton--Hooke Lie algebra in two-dimensional space are given by
\begin{gather*}
[J,K_i]=K_j\epsilon^j_i ,\qquad [J,P_i]=P_j\epsilon^j_i, \\
[K_i,P_j]=M\delta_{ij} ,\qquad [K_i,E]=P_i ,\qquad [P_i,E]=\omega^2
K_i,
\end{gather*}
which means that the generators of space translations as well as
pure Newton--Hooke transformations commute.  One can not then
associate a noncommutative phase space to the Newton--Hooke group. It
is then the absence of the symmetry rotations (anisotropy of the
plane) which guaranties the noncommutative phase space for the
anisotropic Newton--Hooke group.

\subsection{Three-dimensional spaces case}\label{section4.3}

Let $mM^*+h_iJ^{*i}+k_iK^{*i}+p_iP^{*i}+eE^*$, $i=1,2,3$ be the
general element of the dual of the central extended Lie algebra.
Then $m$ and $h_i$ are trivial invariants under the coadjoint action
of $ANH_{\pm}$ in three-dimensional spaces.  We need another
invariant.  As we can verify, the Kirillov's form, in the basis
$(K_i,P_i,E)$, is given by
\begin{gather*}
B_{\alpha\beta}=\begin{pmatrix}
\frac{h_k\epsilon^k_{ij}}{c^2}&m\delta_{ij}&p_i\vspace{1mm}\\-m\delta_{ij}&\pm\frac{h_k\epsilon^k_{ij}}{r^2}&\pm\omega^2k_i\vspace{1mm}\\p_j&\mp\omega^2k_j&0
\end{pmatrix}.
\end{gather*}
The other invariant which is a solution of the
Kirillov's system is
\begin{gather*}
U=e-\frac{p_ip_j\big(\Phi_{\pm}^{-1}\big)^{ij}}{2m}-\frac{m\omega^2q^iq^j\big(\Phi_{\pm}^{-1}\big)_{ij}}{2}+\omega^2p_iq^j\big(\Phi_{\pm}^{-1}A\big)^i_j,
\end{gather*}
where
\begin{gather*}
A_{ij}=\frac{h_k\epsilon^k_{ij}}{mc^2} ,\qquad \Phi_{\pm}=I\pm \omega^2
A \qquad \mbox{and} \qquad  q_i=\frac{k_i}{m}.
\end{gather*}
We see that $\Phi_{\pm}$ is a metric for ${\mathbb R}^3$.  Let us denote the
maximal coadjoint orbit by ${\cal}{O}_{(m,\vec{h},U)}$.  The
restriction of the Kirillov's form on the orbit is then
\begin{gather*}
\Omega=m\begin{pmatrix}
A_{ij}&\delta_i^j\vspace{1mm}\\-\delta^i_j&\pm\omega^2A_{ij}
\end{pmatrix}
\end{gather*}
and its inverse is
\begin{gather*}
\Omega^{-1}=\frac{1}{m}\begin{pmatrix}
\pm
\omega^2\big(A\Phi_{\pm}^{-1}\big)_{ij}&\big(\Phi_{\pm}^{-1}\big)^j_i\vspace{1mm} \\-\big(\Phi_{\pm}^{-1}\big)^i_j&\big(A\Phi_{\pm}^{-1}\big)^{ij}
\end{pmatrix}.
\end{gather*}
The maximal orbit is then equipped with the symplectic structure
\begin{gather*}
\sigma=\big(\Phi_{\pm}^{-1}\big)^i_jdp_i\wedge
dq^j+\frac{1}{m}\big(A\Phi_{\pm}^{-1}\big)^{ij}dp_i\wedge dp_j\pm
m\omega^2\big(A\Phi_{\pm}^{-1}\big)_{ij}dq^i\wedge dq^j
\end{gather*}
and it follows that the Poisson brackets of two functions def\/ined on
the orbit is then
\begin{gather*}
\{f,g\}=(\Phi_{\pm}^{-1})^j_i\left(\frac{\partial f}{\partial
p_i}\frac{\partial g}{\partial q^j}-\frac{\partial f}{\partial
q^i}\frac{\partial g}{\partial p_j}\right)+F_{ij}\frac{\partial
f}{\partial p_i}\frac{\partial g}{\partial p_j}+G^{ij}\frac{\partial
f}{\partial q^i}\frac{\partial g}{\partial q^j}.
\end{gather*}
This implies that
\begin{gather*}
\{p_i,p_j\}=F_{ij} ,\qquad \{p_i,q^j\}=\big(\Phi_{\pm}^{-1}\big)^j_i , \qquad \{q^i,q^j\}=G^{ij},
\end{gather*}
where the magnetic f\/ield $F_{ij}$ and the dual magnetic f\/ield
$G^{ij}$ are given by
\begin{gather*}
 F_{ij}=\pm m \omega^2\big(A\Phi_{\pm}^{-1}\big)_{ij}\qquad \mbox{and}\qquad   G^{ij}=\frac{1}{m}\big(A\Phi_{\pm}^{-1}\big)^{ij}.
\end{gather*}
Moreover the Hamilton's equations are
\begin{gather*}
\frac{dp_k}{dt} = -\big(\Phi^{-1}_{\pm}\big)^i_k\frac{\partial H}{\partial
q^i}\pm m\omega^2\big(A\Phi_{\pm}^{-1}\big)_{ik}\frac{\partial H}{\partial
p_i},\\
\frac{dq^k}{dt} = \big(\Phi^{-1}_{\pm}\big)^k_i\frac{\partial
H}{\partial p_i}+\frac{1}{m}\big(A\Phi^{-1}_{\pm}\big)^{ik}\frac{\partial
H}{\partial q^i}.
\end{gather*}
With the anisotropic Newton--Hooke groups $ANH_{\pm}$ in three-dimensional spaces, we also have realized  phase
spaces where the momenta as well as the positions do not commute.

\section{Conclusion}\label{section5}

We know that we can introduce the classical electromagnetic
interaction through the modif\/ied symplectic form $\sigma=dp_i\wedge
dq^i+\frac{1}{2}F_{ij}dq^i\wedge dq^j$
\cite{abraham,guillemin,souriau}.  This has been
initiated by J.M.~Souriau~\cite{souriau} in the seventies. Recently
many authors (see, e.g.,~\cite{horvathy11,horvathy12,vanhecke11,vanhecke2}) generalized this modif\/ication of
the symplectic form by introducing the so-called dual
magnetic f\/ield such that $\sigma=dp_i\wedge
dq^i+\frac{1}{2}F_{ij}dq^i\wedge dq^j+\frac{1}{2}G^{ij}dp_i\wedge
dp_j$.  The f\/ields $F$ and $G$ are responsible of the
noncommutativity respectively of momenta and positions.  In our paper
we have introduced these f\/ields f\/irstly by minimal coupling momenta  with magnetic potentials (the usual one), secondly by minimal
coupling of positions with dual potentials and lastly by mixing
the two couplings. We have also realized phase spaces endowed with modif\/ied
symplectic structures as coadjoint orbits of the anisotropic Newton--Hooke groups in two
and three-dimensional spaces.  In all these cases, the f\/ields are
constant because they are coming from central extensions of Lie algebras.

\pdfbookmark[1]{References}{ref}
\LastPageEnding


\begin{thebibliography}{99}

\footnotesize\itemsep=0pt

\bibitem{abraham}
Abraham R., Marsden J.E.,
Foundations of mechanics, 2nd ed.,
 Benjamin/Cummings Publishing Co., Inc., Advanced Book Program, Reading, Mass., 1978.


\bibitem{derome}
Derome J.R., Dubois J.G.,
Hooke's symmetries and nonrelativistic cosmological kinematics.~I,
\href{http://dx.doi.org/10.1007/BF02734453}{{\it Nuovo Cimento~B}} {\bf 9} (1972), 351--376.

\bibitem{horvathy11}
Duval C., Horv\'athy P.A.,
The exotic Galilei group and the ``Peierls substitution'',
\href{http://dx.doi.org/10.1016/S0370-2693(00)00341-5}{{\it Phys. Lett.~B}} {\bf 479} (2000), 284--290,
\href{http://arxiv.org/abs/hep-th/0002233}{hep-th/0002233}.

\bibitem{horvathy12}
Duval C., Horv\'athy P.A.,
Exotic Galilean symmetry in the non-commutative plane and the Hall ef\/fect,
\href{http://dx.doi.org/10.1088/0305-4470/34/47/314}{{\it J.~Phys.~A: Math. Gen.}} 34 (2001), 10097--10107,
\href{http://arxiv.org/abs/hep-th/0106089}{hep-th/0106089}.


\bibitem{horvathy13}
Duval C., Horv\'ath Z., Horv\'athy P.A.,
Exotic  plasma as classical Hall liquid,
\href{http://dx.doi.org/10.1142/S0217979201007361}{{\it Internat.~J. Modern Phys.~B}} {\bf 15} (2001),  3397--3408,
\href{http://arxiv.org/abs/cond-mat/0101449}{cond-mat/0101449}.

\bibitem{grigore}
Grigore D.R.,
Transitive symplectic manifolds in $1+2$ dimensions,
\href{http://dx.doi.org/10.1063/1.531388}{{\it J.~Math. Phys.}} {\bf 37} (1996), 240--253.

\bibitem{guillemin}
Guillemin V., Sternberg  S.,
Symplectic techniques in physics,
Cambridge University Press, Cambridge, 1984.

\bibitem{hamermesh}
 Hamermesh M.,
 Group theory and its applications to physical problems,
{\it Addison-Wesley Series in Physics},  Addison-Wesley Publishing Co., Inc., Reading, Mass.~-- London, 1962.

\bibitem{horvathy2}
Horv\'athy P.A.,
The non-commutative Landau problem,
\href{http://dx.doi.org/10.1006/aphy.2002.6271}{{\it Ann. Physics}} {\bf 299} (2002), 128--140,
\href{http://arxiv.org/abs/hep-th/0201007}{hep-th/0201007}.

\bibitem{kirillov}
Kirillov A.A.,
Elements of theory of representations,
{\it Grundlehren der Mathematischen Wissenschaften}, Band~220, Springer-Verlag, Berlin~-- New York, 1976.

\bibitem{kostant}
Kostant~B.,
Quantization and unitary representations. I.~Prequantization,
in Lectures in Modern Analysis and Applications,~III,
{\it  Lecture Notes in Math.}, Vol.~170, Springer, Berlin, 1970, 87--208.

\bibitem{nzo1}
Nzotungicimpaye J.,
Galilei--Newton law by group theoretical methods,
\href{http://dx.doi.org/10.1007/BF00397830}{{\it Lett. Math. Phys.}} {\bf 15} (1988), 101--110.

 \bibitem{nzo2}
 Nzotungicimpaye J.,
 Jerk by group theoretical methods,
\href{http://dx.doi.org/10.1088/0305-4470/27/13/025}{{\it J.~Phys.~A: Math. Gen.}} {\bf 27} (1994), 4519--4526.

\bibitem{peierls}
Peierls R.,
On the theory of diamagnetism of conduction electrons,
{\it Z.~Phys.} {\bf 80} (1933), 763--791.

\bibitem{romero}
Romero J.M., Santiago J.A., Vergara J.D.,
Newton's second law in a non-commutative space,
\href{http://dx.doi.org/10.1016/S0375-9601(03)00191-9}{{\it Phys. Lett.~A}} {\bf 310} (2003), 9--12,
\href{http://arxiv.org/abs/hep-th/0211165}{hep-th/0211165}.

\bibitem{snyder}
Snyder H.S.,
Quantized space-time,
\href{http://dx.doi.org/10.1103/PhysRev.71.38}{{\it Phys. Rev.}} {\bf 71} (1947), 38--41.

\bibitem{souriau}
Souriau J.-M.,
Structure des syst\`emes dynamiques,
Ma\^{\i}trises de Math\'ematiques Dunod, Paris, 1970.

 \bibitem{vanhecke11}
 Vanhecke F.J., Sigaud C., da Silva A.R.,
Noncommutative conf\/iguration space. Classical and quantum mechanical aspects,
 {\it Braz.~J. Phys.} {\bf 36} (2006), 194--207,
\href{http://arxiv.org/abs/math-ph/0502003}{math-ph/0502003}.

\bibitem{vanhecke2}
Vanhecke F.J., Sigaud C., da Silva A.R.,
Modif\/ied symplectic structures in cotangent bundles of Lie groups aspects,
 {\it Braz.~J. Phys.} {\bf 39} (2009), 18--24,
\href{http://arxiv.org/abs/0804.1251}{arXiv:0804.1251}.

\bibitem{wei}
Wei G.-F., Long C.-Y., Long Z.-W., Qin S.-J., Fu Q.,
Classical mechanics in non-commutative phase space,
\href{http://dx.doi.org/10.1088/1674-1137/32/5/002}{{\it Chinese Phys.~C}} {\bf 32} (2008), 338--341.

\end{thebibliography}
\end{document}